\documentclass[showpacs,aps,pra,twocolumn,amsmath,amssymb,superscriptaddress]{revtex4-2}
\usepackage{amsmath}           
\usepackage{bbm}
\usepackage{wasysym}
\usepackage{tikz}
\usetikzlibrary{arrows.meta,calc,positioning}
\usepackage{color}           
\usepackage[latin1]{inputenc}  
\usepackage{graphicx}
\begin{document}
\title{On quantum mechanics self-consistency: EPR incompleteness
claims require no extraneous concepts beyond the theory's plain
formalism for refutation
}
\author{D. F. Orsini, L. R. N. Oliveira, M. G. E. da Luz}
\affiliation{Departamento de F\'{\i}sica
  \&
Multidisciplinary Laboratory for Modeling and Analysis
of Data in Complex Systems (MADComplex), N\'ucleo de 
Modelagem e Computa\c c\~ao Cient\'{\i}fica-Centro 
Interdisciplinar de Ci\^encia, Tecnologia e 
Inova\c c\~ao, Universidade Federal do
  Paran\'a, Curitiba-PR, 81531-980, Brazil}
\email{luz@fisica.ufpr.br}

\date{\today}

\begin{abstract}
From a phenomenological and experimental viewpoint, quantum
mechanics is remarkably successful in explaining 
effects and processes in the microscopic world.
Recently, however, a growing body of literature has revisited
the theory's own formal foundations, specifically addressing
issues of self-consistency and completeness.
As a major historical example of these foundational challenges,
the Einstein-Podolsky-Rosen (EPR) argument famously claimed that
the theory is incomplete.
While many valid criticisms and refutations of the EPR logic
exist, they generally rely on a combination of technical results
and conceptual objections to EPR's interpretive assumptions,
thus transcending the plain quantum framework itself.
Motivated by these trends of examining the theory's structural
limits, here we revisit the EPR argument.
Focusing on elements essential to the EPR reasoning, we first
analyze general quantum correlations for EPR states:
(i) the fact that their observables are always associated with
non-commuting operators, and (ii) the nature of the correlated
information obtained through measurement.
From these two points alone, and relying strictly on the core
rules of quantum mechanics, we demonstrate how to overturn the
alleged EPR incompleteness without invoking any extraneous
propositions.
Consequently, we show that the standard formalism alone suffices
to resolve such type of skepticism regarding the theory's
physical reach.
This offers, at least in a paradigmatic instance, a powerful
indication of a structurally well-founded, self-consistent
quantum theory.
\end{abstract}


\maketitle


\section{Introduction}
Frequently, the already centennial framework of quantum mechanics (QM)
is stated to be the most successful theory in all of physics.
Indeed, QM's remarkable power in describing the microscopic realm,
leading to countless technological applications, allied with a vast
body of accumulated empirical evidence testing and confirming
its foundations and predictions, has established the theory
as a pillar of modern scientific knowledge.
But parallel to steady technical and conceptual developments,
structural questions associated with the foundational
\cite{tisza-1989,benioff-1999,ferrero-2013,hordecki-2015} and
universal \cite{collins-1970,deutsch-1985,branford-2018} aspects
of QM have naturally emerged.
Queries regarding its reach of validity and generality
\cite{collins-1970,fine-1974,colbeck-2011}, relation to causality
\cite{allen-2017,harper-2017,jia-2018}, nonlocality and realism
\cite{leggett-2003,horodecki-2009,popescu-2014}, and even
difficulties with self-reference \cite{benioff-1999,frauchiger-2018},
have been discussed under different perspectives.

To a great extent, these works aim to address the
self-consistency and completeness \cite{fine-1974,colbeck-2011}
of QM, validating it as a proper representation of nature.
Assuming that QM is (within its domain) capable of explaining any
physical phenomenon --- though, as a high-level systematic
theory, it is not immune to the limitations imposed by G\"odel's
incompleteness theorems, as illustrated by the potential
undecidability of the spectral gap problem \cite{cubitt-2015} ---
a key question is how straightforwardly one can portray a
process based solely on the formal mathematical construction of
its basic laws and assumptions, namely, without invoking
propositions outside its common core (such as additional
characterizations of physical reality).
If QM proves to be such a self-contained functional theory,
one can conceivably ascribe solid ontological significance to
many of its key ideas, as already established for the quantum
state \cite{harrigan-2010,pusey-2012,mansfield-2016,myrvold-2018}.
Hence, QM would directly yield a comprehensive picture of the
empirical world without relying on extra metaphysical concepts
\cite{mermin-1990}.
As clearly put in \cite{stanford-web-interpreation}:
``One should not be misled by \ldots \ thinking that what we have
is an uninterpreted mathematical formalism [of QM]
with no connection to the physical world.
Rather, there is a common core of interpretation that consists
of recipes for calculating probabilities of outcomes of
experiments performed on systems subjected to certain state
preparation procedures.
[But what] \ldots \ are often referred to as different
interpretations of quantum mechanics differ on what,
if anything, is added to the common core.
Arguably, two of the major approaches, hidden-variables
theories and collapse theories, involve formulation of physical
theories distinct from standard quantum mechanics \ldots''.

Given such a scenario of analyzing the actual scope of QM, one
can ask which concrete physical problems can be settled solely from
its laws and principles \cite{stanford-web-interpreation};
in other words, when interpretative or epistemological arguments
are unnecessary to prove the truth of a QM statement about an
objective event \cite{espagnat-1979}.
While tackling this in general terms remains challenging,
particular (emblematic) situations can shed light on the completeness and
self-evidence of QM.
In this contribution, we provide an important instance where the
standard rules of QM alone refute allegations of inconsistency.
We do so by revisiting the Einstein-Podolsky-Rosen (EPR) logic,
which famously argued that QM is incomplete \cite{epr}.
The EPR paper \cite{epr} is certainly one of the most stimulating
in the history of physics.
Dissected and re-examined for over ninety years, it has spurred
foundational advances such as entanglement, nonlocality, and
Bell's inequalities and breakthrough experiments
\cite{aspect-1982-1,aspect-1982-2}.

Although a large body of literature opposes
the EPR conclusion via distinct valid objections ---
ranging from criticisms of EPR's definitions of physical
reality and local causality to the use of counterfactual
inference (see \cite{stanford-web} and references therein) ---
these refutations often transcend operational QM, entering an
interpretative debate \cite{wiseman-2006}.
Even the assumptions used to derive Bell's inequalities and
demonstrate their violation by QM \cite{bell-1964,bell-1986}
are based on conceptions foreign to the bare quantum
formalism (see, e.g., \cite{blaylock-2010}).
Hence, we shall show that the apparent inner paradox of the EPR
argument is fully resolved by the structural consistency of standard
QM.
We note that while simpler versions of the EPR thought-experiment
exist --- e.g., those avoiding distinct observables associated
with non-commuting operators --- such contexts pose greater
difficulties for a rigorous probing \cite{wiseman-2006}.
Thus, we focus exclusively on the traditional scenario of
\cite{epr}.

The line of argumentation in \cite{epr} is to: (1a) analyze a
system described by an EPR state $| \Psi \rangle$; (1b) assume
the formal rules of QM governing $| \Psi \rangle$ are valid;
and (2) consider two extra assumptions foreign to the conventional
framework of QM, namely, the notion of an ``element of physical
reality'' and a statement of non-disturbance between systems
(premises detailed below).
The EPR conclusion is that, within the breadth of QM, (1)--(2)
result in pitfalls preventing a complete description of a system
by a quantum state.
For such a proof scheme to be logically consistent, however, (3)
any presumed physical process must verify (1b); otherwise, one
would not be properly handling the theory under scrutiny.
From the plain rules of QM, we examine two specificities of EPR
states $\vert{}\Psi\rangle$: (i) a particular quantum correlation
valid for any $\vert{} \Psi \rangle$ (and explicitly manifested in
traditional examples involving position--momentum, photon
polarization, or spin-1/2 components); and (ii) the nature of
the correlated information obtained by measuring a
given $\vert{}\Psi\rangle$.
Taking these features into account within the EPR rationale
conforming to (3), we find they overturn the alleged QM
inconsistency in \cite{epr}.

The letter is organized as follows.
We first outline the key properties of the state $\vert{}\Psi \rangle$
that are fundamental to the EPR rationale.
Next, we recall established facts regarding the core correlation
(item (i) above) of the associated observables, namely, that
they must correspond to non-commuting operators.
We then revisit the EPR argument in light of points (i) and (ii),
showing that it can be refuted exclusively from the QM formalism.
Finally, we briefly highlight a fundamental conceptual
perspective of our analys, the QM mathematical treatment of single
versus composite systems.

\section{The pairwise incompatibility of observables in EPR states}
The EPR reasoning, aimed at showing that QM is incomplete,
relies on portraying a system via a specific entangled state
$\vert{}\Psi \rangle$.
According to EPR, this leads to the conclusion that \cite{epr}
``the description of reality [of such a system] as given by a wave
function [the EPR state $\vert{}\Psi \rangle$] is not complete''.
Before summarizing the full EPR argument, we focus on the required
features of these states.
For convenience, we adopt a notation slightly different from
\cite{epr} but strictly faithful to its original ideas.
Although a formal construction is presented in \cite{epr} (including
examples), explicit assumptions and general
mathematical properties of $\vert{}\Psi \rangle$ were not
fully addressed.
Indeed, \cite{epr} was published in 1935, when landmark works
developing the mathematical foundations of QM --- such as those by
von Neumann \cite{neumann-1929,neumann-1932,neumann-1934,neumann-1936}
--- were only beginning to establish today's rigorous formal
structures.

The preparation details of the EPR state are irrelevant here.
Crucially, two systems, I and II, interact to correlate certain
``physical quantities'' of I (say, $C$, associated with the
Hermitian operator $\hat{C}$) with those ``pertaining'' to II (say,
$A$, associated with $\hat{A}$), yielding the entangled state
\cite{schoredinger-1935}
$\vert{}\Psi \rangle = \sum_n \vert{}c_n \rangle_{I} \otimes
\vert{}a_n \rangle_{II}$ (hereafter global normalizatoin constant
will be omitted).
Afterwards, the systems are carefully separated to preserve this
entanglement, nevertheless such that ``$\ldots$ there is no longer
any interaction between the two parts''.
This spatially disengaged state is assumed to persist until proper
measurements.
A second fundamental supposition, condition (b), allows
$\vert\Psi\rangle$ to be written in terms of alternative observables
$D$ (for I) and $B$ (for II), such that
\begin{equation}
  |\Psi \rangle = \sum_n \, |c_n \rangle_{I} \otimes
  |a_n \rangle_{II}= \sum_n \,
  |d_n \rangle_{I} \otimes |b_n \rangle_{II}.
  \label{eq:exp-a}
\end{equation}
Reference \cite{epr} illustrates the feasibility of
Eq.~(\ref{eq:exp-a}) using two 1D particles for the observables
being positions and momenta, but then develop their
arguments for the arbitrary case in Eq. (\ref{eq:exp-a}).
For EPR type of analysis, one can assume that any
term $|e_n \rangle_{I} \otimes |f_n \rangle_{II}$
has the same complex amplitude \cite{arens-2000,reid-2009},
exactly as done in \cite{epr}.
Fundamental to the EPR rationale is ensuring that a measurement
on I alone --- determining either $C$ or $D$ --- provides full
knowledge about the corresponding value of $A$ or $B$ for II,
a result which can be further verified upon subsequent inspection
of system II if the same observable ($A$ or $B$) is probed.
This requirement defines condition (c).
Lastly, condition (d) (as stipulated by EPR) demands that the
observables $A$ and $B$ of system II correspond to non-commuting
operators, $[\hat A, \hat B] \neq 0$.

Notably, \cite{epr} imposes no explicit conditions on the observables
$C$ and $D$ of system I.
A legitimate question then arises regarding what further
characteristics $|\Psi\rangle$ must possess to comply with (a)--(d).
This is no mere technicality; it lies at the core of the EPR debate.
Indeed, \cite{epr} argues that quantum mechanics is incomplete by
claiming that ``predictions concerning a system [described by an
EPR state] on the basis of measurements'' yield more information
than allowed by $|\Psi\rangle$.
Nonetheless, EPR assume that their proposed experiments and expected
outcomes are fully compatible with the intrinsic properties of
$|\Psi\rangle$.
A fundamental fact completely overlooked by EPR is that any expansion
in the form of Eq.~(\ref{eq:exp-a}) with $[\hat A, \hat B] \neq 0$
for system II strictly requires the corresponding observables
of system I to be non-commuting as well, mandating that
$[\hat C, \hat D] \neq 0$ without exception.
Curiously, while such a property was implicitly present in specific
contexts, as Bohr's historic reply \cite{bohr-1935} to EPR
and Bohm's spin-1/2 model \cite{bohm-book}, this essential and
general necessity was pinpointed and rigorously demonstrated in
full only recently \cite{orsini-2024}.

\section{The EPR original incompleteness argument}
For convenience, the relevant assumptions from \cite{epr} are
explicitly quoted and labeled as S1, S2, \ldots (a notation not
present in the original text) to be referenced hereafter.
EPR starts with a broad view on ``physical reality'', relating it
to the formal structure of QM.
In particular, they introduce a pragmatic ``necessary'' criterion
for a theory to be complete (S1):
``{$\ldots$ every element of the physical reality must have a
  counterpart in the physical theory}''.
Then, EPR propose a ``comprehensive'' definition of elements of
reality --- regarded as a ``sufficient'' condition --- assuming
that (S2):
``{If, without in any way disturbing a system, we can predict
with certainty (i.e., with probability equal to unity) the value of a
physical quantity, then there exists an element of physical reality
corresponding to this physical quantity}''.
Further in \cite{epr}, it is admitted that, as far as quantum theory
is concerned, the description of a system's behavior is ``completely
characterized'' by the concept of a ``{state}'' (or wave function).
Moreover, if the system state $|\psi \rangle$ is an eigenstate of a
Hermitian operator $\hat A$ with eigenvalue $a$
($\hat A |\psi \rangle = a |\psi \rangle$), then:
``$\ldots$ there is an element of physical reality corresponding
to the physical quantity $A$''.

In QM, if the respective operators for two observables of the
same system do not commute, $[\hat A, \hat B] \neq 0$, then (S3):
``the precise knowledge of one of them precludes such a knowledge
of the other \ldots \ any attempt to determine the latter
experimentally will alter the state of the system in such a way as
to destroy the knowledge of the first''.
This statement is fundamental, as EPR implicitly assume that their
reasoning does not violate S3.
Very importantly, as we shall see, this is not the case.

Given S2 and $[\hat A, \hat B] \neq 0$, whenever a state
$| \psi \rangle$ ascribes reality to $A$ ($B$), it cannot also
ascribe reality to $B$ ($A$).
But if objectively both $A$ and $B$ can possess
``simultaneous reality'', then according to S1, the description
given by $\vert{} \psi \rangle$ is incomplete \cite{epr}.
Such arguments lead to the alternative conclusion S4:
``{Either (1) the quantum-mechanical description of reality given
  by the wave function is not complete or (2) when the operators
  corresponding to two physical quantities do not commute the
  two quantities cannot have simultaneous reality}''.

\begin{figure}[!t]
\includegraphics[width=1\linewidth]{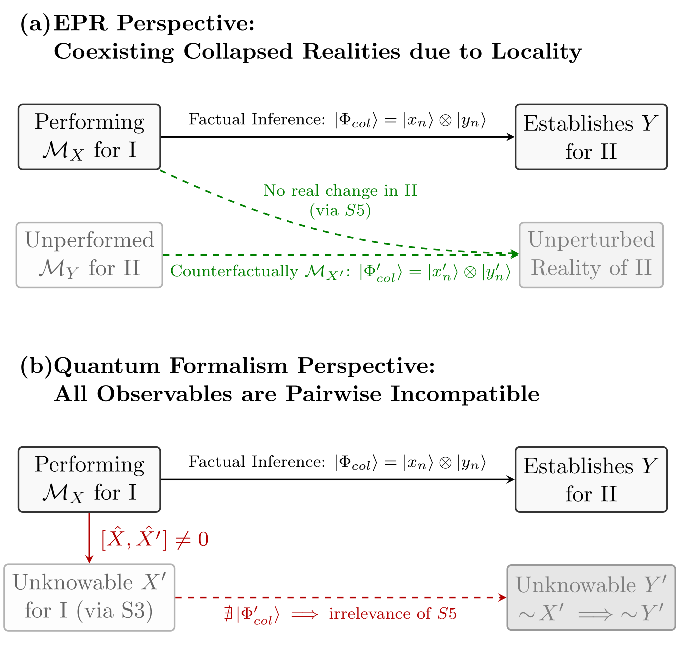}
\caption{
For an EPR state, a measurement of $X \in \{C, D\}$ (with
$X'\in \{D, C\}$) is performed on I, while nothing is done
to II (where $Y \in \{A, B\}$ and $Y' \in \{B, A\}$).
Within this setup:
(a) Relying on $[\hat{Y}, \hat{Y}'] \neq 0$ for II while ignoring
$[\hat{X}, \hat{X}'] \neq 0$ for I, the EPR reasoning (here
schematically) claims that QM is incomplete since
$|\Phi_{col} \rangle \neq |\Phi_{col}' \rangle$ coexist due
to S6.
(b) Conversely, quantum formalism alone---leveraging the pairwise
incompatibility of observables in EPR states---overturns these
claims without extraneous assumptions, e.g., in the present
context the validity or not of $S5$ is irrelevant
(main text).
}
\label{fig:fig1}
\end{figure}

From these considerations, EPR proposes a framework (Fig.
\ref{fig:fig1}(a)) from which one should conclude that QM
is incomplete.
Because the entanglement of I and II, the unitary time
evolution of the EPR state $| \Psi \rangle$ alone cannot
associate independent elements of reality (S2) to either
system.
As stated in \cite{epr}, it
``$\ldots$ can be done only with the help of further measurements,
by a process known as the reduction of the wave packet''.
In fact, one of the expansions for $\vert{} \Psi \rangle$ in
Eq.~(\ref{eq:exp-a}) can be probed through the measurement of an
observable, say $C$ or $D$, but not both.

Evoking condition (c), choosing to measure $C$ on system I
(${\mathcal M}_C$) and obtaining $c_n$ ensures that the value
of $A$ for system II must be $a_n$, leaving the collapsed state
as $\vert{}c_n \rangle_I \otimes \vert{}a_n \rangle_{II}$.
By the same token, determining $D$ (${\mathcal M}_D$) and finding
$d_n$ implies that the value of $B$ for II is $b_n$, yielding the
state $\vert{}d_n \rangle_I \otimes \vert{}b_n \rangle_{II}$.
Hence, the choice of measurement on system I imposes --- according
to S2 --- a specific element of reality for system II,
unknown prior to the measurement.

EPR then propose what non-interacting systems ultimately should
convey (S5): ``$\ldots$ since at the time of measurement the two
systems no longer interact, no real change can take place in the
second system in consequence of anything that may be done to the
first system.
This is, of course, merely a statement of what is meant by the
absence of an interaction between the two systems''.
S5 is highly controversial from a modern perspective
\cite{alford-2016}, constituting the main point of debate for
later rejections of the EPR conclusions, including Bohr's
reply \cite{bohr-1935}.

Next, EPR assert the implications of S5 on inferences regarding
$| \Psi \rangle$.
While QM forbids a definite reality for both
$A$ and $B$ if the measurements are performed directly on
II (see S3), \cite{epr} notes: ``as a consequence of two different
measurements performed upon the first system [I], the second
system [II] may be left in state with two different wave
functions [namely, $\vert{} a_n \rangle_{II}$ in ${\mathcal M}_C$
and $\vert{} b_n \rangle_{II}$ in ${\mathcal M}_D$]''.
But because S5 demands that actions on I cannot influence II,
EPR state that ``no real change can take place'' in II.
Therefore, 
S6: ``{\ldots it is possible to assign two different wave functions
  \ldots to the same reality}'', each describing a concrete value
for one of the two non-commuting observables.
Thus, both $A$ and $B$ for II would constitute simultaneous
elements of reality.

However, according to QM, the final collapsed state must be
unique, fully describing the composite system.
The wave function characterization of the outcomes
--- $\vert{}c_n \rangle_I \otimes \vert{}a_n \rangle_{II}$
(not an eigenstate of $\hat B$) in one case and $\vert{}d_n
\rangle_I \otimes \vert{}b_n \rangle_{II}$ (not an eigenstate
of $\hat A$) in the other --- is incompatible with the EPR
alleged concurrent reality of $A$ and $B$.

Finally, the last step in the EPR reasoning confronts this
with statement S4.
Although the logic behind this step has been questioned
\cite{brassard-2006,brassard-2010}, EPR argue that the negation
of alternative (2) in S4 --- namely, that two physical quantities
with non-commuting operators can have simultaneous reality ---
forces the truth of alternative (1):
``We are thus forced to conclude that the quantum-mechanical
description of physical reality given by wave functions is not
complete''.
This whole line of thought is summarized in Fig. \ref{fig:fig1}(a).

As a closing comment, EPR digress about potential criticisms
``\ldots on the grounds that our criterion of reality is not
sufficiently restrictive'', a point that forms the basis of
many rebuttals \cite{plotnitsky-2009}.
Since our analysis strikes in a completely different direction,
this specific debate is not relevant for our purposes here.

\section{Previous limited scope incompatibility-based refutations}
As already mentioned, paramount to the EPR logic are premises
(1a), (1b), (2), and (3).
Criticisms of EPR typically challenge (2) rather than (3).
For instance, historical refutations --- such as Bohr's response
\cite{bohr-1935} and Bohm's spin-$1/2$ reformulation
\cite{bohm-book} --- do not address the general incompatibility
of $C$ and $D$ for system I, treating it as a secondary technical
detail rather than a universal constraint.
Examining position and momentum, Bohr argued that a
setup measuring $C$ or $D$ on I precludes the simultaneous
reality of $A$ and $B$ envisioned by EPR, as apparatuses
suited for $A$ fail for $B$ due to the uncertainty principle.
From such case-dependent scenarios, Bohr attempted to
generalize, although not without controversy \cite{fine-2007},
that S2 and S5 conflict with actual experiments.
In fact, Bohr's tabletop apparatus for position and momentum,
based on interconnected parts and leading to the limitation of the
``impossibility... of accurately controlling the reaction of the
object on the measuring instruments'' \cite{bohr-1935} is clearly
dated.
Modern experiments measuring electron spins $1.3\text{km}$
apart \cite{hensen-2015} or rubidium atoms separated by $398\text{m}$
\cite{rosenfeld-2017} transcend this device-specific complementarity
dismissing of EPR.
However, other more elaborate objections
\cite{halvorson-2002,bacciagaluppi-2015} similarly rely on
specific measurement implementations.

In broader terms, a core difficulty with
unspecified observables $C$ and $D$ is that actual measurements
on system I cannot always be explicitly devised, leaving 
concrete conditions that hinder a unique device for $A$ or $B$
unpinpointed.
For genuine robustness, QM should provide a universal
argument invalidating the EPR conclusion independently of the
particular features of $C$ and $D$.
Moreover, theoretical self-consistency requires quantum predictions
to hold without relying on hypotheses extraneous to the core
formalism.
We next demonstrate that $[\hat C, \hat D] \neq 0$, combined
with the information extractable from an EPR state, supplies
this universal argument.
Fundamentally, this approach avoids disputes over the validity
of S2 and S5 or the need to reject EPR counterfactual reasoning
(which generally clashes with QM, though see \cite{laudisa-2018}).

\section{Refuting EPR incompleteness purely from quantum formalism}
Before presenting our key result, we note existing robust
objections to strict notions of the EPR ``element of physical
reality'' \cite{mermin-1990,greenberger-1989,groblacher-2007}.
For instance, physical reality can become ambiguous when observables
exhibit quantum correlations with an agent recording state
information \cite{bilobran-2015,gomes-2018}.
Here, we retain the concept (along with S2) while avoiding
conflicts with the quantum standard theory by treating it
strictly as a label ---
a characterization of a system that, following a suitable physical
process, possesses an observable predictable with $100\%$
confidence via proper measurements.
Presumably, this mirrors the intended connotation in \cite{epr}.

For a composite system described by $|\Psi\rangle$, EPR concede
the inability to ``calculate the state in which either one of the
two systems is left after the interaction.''
Under S2 and standard premises --- as assumed in \cite{epr} ---
one must assume no pre-existing elements of reality for
$C\text{--}A$ or $D\text{--}B$ prior to measuring
$|\Psi\rangle$.
However, inspecting $C$ ($D$) on system I specifies the value of
$A$ ($B$) on system II (EPR condition (c)), yielding joint knowledge
of one quantity per system.
Further, recalling the view in \cite{epr} (S7) that ``elements
of physical reality cannot be determined by {\em a priori}
philosophical considerations [but] must be found by an appeal
to results of experiments and measurements''.
Consequently, elements of physical reality emerge either
(i) for both systems or (ii) for neither, as a direct result of
wave-packet reduction \cite{epr}.
Furthermore, outcome (i) does not conflict with ascribing reality
to each system individually.
Indeed, after the collapse of $|\Psi\rangle$, any subsequent
process on II (I) depends solely on II (I) having become an eigenstate
of $A$ or $B$ ($C$ or $D$), independently of the other subsystem.
Importantly, these considerations carry concrete physical
significance rather than mere semantics \cite{fine-2007}.

Although the elements of physical reality for systems I and II
(defined by S2 and verified via S7) are autonomous, their
{\em mutual existence} is strictly correlated.
Information obtainable, by probing $|\Psi\rangle$, about
system I constrains what can be learned about system II:
knowing $A$ for II requires knowing $C$ for I (and similarly
for $B$ and $D$).
Ref. \cite{epr} does not deny this fundamental feature (cf. S7);
rather, it is tacit in their argument for incompleteness.
Thus, the emergence of an element of reality for I is inextricably
linked to that for II.
Inspecting $|\Psi\rangle$ via measurements of $C$ or $D$ yields a
joint outcome characterized strictly by one of two joint
probability distributions or paired states:
\begin{equation}
  P(C) \equiv p(c_n|_I, a_n|_{II}) \quad \text{or} \quad P(D)
  \equiv p(d_n|_I, b_n|_{II}),
\label{eq:comp-real}
\end{equation}
where $(x|_I, y|_{II})$ designates the joint inference of $x$
for system I and $y$ for system II, Fig. \ref{fig:fig1}.

Independently, these considerations do not invalidate the EPR
logic; they merely describe what standard quantum rules allow
to infer from $|\Psi\rangle$.
Setting aside concrete violations of Bell inequalities for
the sake of argument, the realization of elements of physical
reality in EPR states could conceivably align with two distinct
interpretations:
\begin{itemize}
\item[(A)]
Non-local correlations \cite{bell-1987,aspect-2007} exist
between spatially
separated systems, meaning reality elements are actualized
for both systems upon probing system I, which instantaneously
affects system II.
\item[(B)]
Complying with local realism (S5), $|\Psi\rangle$ encodes only
partial statistical information about non-interacting systems.
This incomplete knowledge, when supplemented by measurements on
system I, unveils a pre-existing element of reality on system II
(potentially via S6).
However, as required by S7, perceiving this pre-existence must
follow experimental analysis rather than {\em a priori}
assumptions --- a principle explicitly invoked in \cite{epr}.
\end{itemize}
The QM rules out interpretation (B).
Crucially, however, the ``complete description of the physical
reality'' of an EPR state (\cite{epr}, p. 2) is entirely captured
by Eq. (\ref{eq:comp-real}), regardless of interpretations
regarding localized state updates on system II, namely, whether
S5 holds.

Now, incorporating $[\hat C, \hat D] \neq 0$:
First, because $\hat A$ and $\hat B$ do not commute on system II,
S3 prohibits simultaneous elements of reality for those observables
if measured directly on II, which motivates the EPR strategy of
probing system I instead.
However, since $[\hat C, \hat D] \neq 0$, consistency with S3
imposes that $C$ and $D$ cannot be concurrent observables for
system I either.
Second, QM dictates that $A$ and $C$ ($B$ and $D$)
form combined elements of physical reality, represented by Eq.
(\ref{eq:comp-real}).
This is explicitly admitted in \cite{epr} (p. 3,
third and fourth paragraphs), forming what
we designate as EPR condition (e).
Since $C$ and $D$ cannot be simultaneous elements of physical
reality, this incompatibility extends to their corresponding joint
realities $P(C)$ and $P(D)$, and consequently to the associated
observables $A$ and $B$, thereby invalidating the main EPR
assertion {\em directly from standard quantum formalism}.
This simple argument is depicted in Fig. \ref{fig:fig1}(b).

As a last observation, relying on S5, EPR claim counterfactually
(Fig. \ref{fig:fig1}(a)) that choosing to measure $C$ or $D$
on system I should not modify system II, leaving its pre-existing
reality unaltered.
Under this assumption, the distinct states $| a_n \rangle_{II}$
and $| b_n \rangle_{II}$ for incompatible observables would
contradictorily belong to ``the same reality'' (S6).
The crucial point is that irrespective of S5's validity,
both $[\hat C, \hat D] \neq 0$ and condition (e) imply that
inferred states $| a_n \rangle_{II}$ and $| b_n \rangle_{II}$
cannot portray the same element of reality, because their
respective generating states on system I ($| c_n \rangle_I$
and $| d_n \rangle_I$) cannot either (Fig. \ref{fig:fig1}(b)).
This fundamental gap in \cite{epr}, as far as we know, has
not being addressed anywhere in the literature.
%
%

\section{Concluding remarks}
The incompatibility of observables in single systems (e.g.,
position and momentum) and multipartite entanglement represent
foundational quantum correlations \cite{fanchini-2017}.
While single-system non-commutativity ($\lbrack \hat{A}, \hat{B}
\rceil \neq 0$) is universally accepted as a constraint on
simultaneous knowledge, extending it to spatially separated
systems has historically sparked debates over locality and
counterfactual definiteness.
However, regardless of how entanglement is generated
\cite{horodecki-2009,bengtsson-2006}, bipartite states
structured as EPR states must always be expressed in bases
with pairwise incompatible observables \cite{orsini-2024}.
Thus, composite systems naturally inherit the algebraic
non-commutativity of single systems.
Any attempt to infer simultaneous physical reality for
non-commuting observables inexorably collides with this
constraint.
Consequently, the EPR incompleteness argument dissolves:
QM prohibits simultaneous reality across composite systems
precisely as single-system incompatibility is accepted without
paradox.
This structural closure avoids splitting interpretations
between single and composite systems, shifting the true
conceptual departure from classical physics not to resolving
EPR (e.g., via negating the incorrect S5), but to why QM treats single
and multipartite systems on the exact same foundational footing
regarding observable incompatibility.

Finally, a sound scientific framework should resolve foundational
questions entirely through its universal principles ---
a self-consistency tested for QM from its inception to our
resolution of the EPR claim.
While our results are fundamentally technical, they directly
align with naturalistic epistemology \cite{smith-2003},
which proposes that questions of scientific knowledge must be
addressed within the scope of the theory's own formal
structure.
Actually, elucidating the nature of physical reality directly
from the mathematical core of QM demonstrates that a complete
physical theory possesses the inherent power to resolve its own
apparent conceptual conundrums.

Research funding is provided by Capes through a PhD scholarship
program (DFO and LRNO) and CNPq through Proj. 307512/2023-1 (MGEL).

\end{document}